%
%
%
%
\def\unredoffs{} \def\redoffs{\voffset=-.40truein\hoffset=-.40truein}
\def\speclscape{}
%
%
%
%
\newbox\leftpage \newdimen\fullhsize \newdimen\hstitle \newdimen\hsbody
\tolerance=1000\hfuzz=2pt
\catcode`\@=11 
\def\bigans{b }
\def\answ{b }
\ifx\answ\bigans\message{(This will come out unreduced.}
\magnification=800\unredoffs\baselineskip=16pt plus 2pt minus 1pt
\hsbody=\hsize \hstitle=\hsize 
\else\message{(This will be reduced.} \let\l@r=L
\magnification=1000\baselineskip=16pt plus 2pt minus 1pt
\vsize=7truein \redoffs
\hstitle=8truein\hsbody=4.75truein\fullhsize=10truein\hsize=\hsbody
\output={\ifnum\pageno=0 
    \shipout\vbox{\speclscape{\hsize\fullhsize\makeheadline}
      \hbox to \fullhsize{\hfill\pagebody\hfill}}\advancepageno
    \else
    \almostshipout{\leftline{\vbox{\pagebody\makefootline}}}\advancepageno
    \fi}
\def\almostshipout#1{\if L\l@r \count1=1 \message{[\the\count0.\the\count1]}
        \global\setbox\leftpage=#1 \global\let\l@r=R
   \else \count1=2
    \shipout\vbox{\speclscape{\hsize\fullhsize\makeheadline}
        \hbox to\fullhsize{\box\leftpage\hfil#1}}  \global\let\l@r=L\fi}
\fi
%
\newcount\yearltd\yearltd=\year\advance\yearltd by -1900

\def\Title#1#2{\nopagenumbers\abstractfont\hsize=\hstitle\rightline{#1}%
\vskip 1in\centerline{\titlefont #2}\abstractfont\vskip
.5in\pageno=0}
%
%

\def\draftmode{\message{ DRAFTMODE }\def\draftdate{{\rm preliminary draft:
\number\month/\number\day/\number\yearltd\ \ \hourmin}}%
\headline={\hfil\draftdate}\writelabels\baselineskip=20pt plus 2pt
minus 2pt
   {\count255=\time\divide\count255 by 60 \xdef\hourmin{\number\count255}
    \multiply\count255 by-60\advance\count255 by\time
    \xdef\hourmin{\hourmin:\ifnum\count255<10 0\fi\the\count255}}}
\def\nolabels{\def\wrlabeL##1{}\def\eqlabeL##1{}\def\reflabeL##1{}}
\def\writelabels{\def\wrlabeL##1{\leavevmode\vadjust{\rlap{\smash%
{\line{{\escapechar=` \hfill\rlap{\sevenrm\hskip.03in\string##1}}}}}}}%
\def\eqlabeL##1{{\escapechar-1\rlap{\sevenrm\hskip.05in\string##1}}}%
\def\reflabeL##1{\noexpand\llap{\noexpand\sevenrm\string\string\string##1}}}
\nolabels
%
\global\newcount\secno \global\secno=0 \global\newcount\meqno
\global\meqno=1
\def\newsec#1{\global\advance\secno by1\message{(\the\secno. #1)}
\global\subsecno=0\eqnres@t\noindent{\bf\the\secno. #1}
\writetoca{{\secsym} {#1}}\par\nobreak\medskip\nobreak}
\def\eqnres@t{\xdef\secsym{\the\secno.}\global\meqno=1\bigbreak\bigskip}
\def\sequentialequations{\def\eqnres@t{\bigbreak}}\xdef\secsym{}
\global\newcount\subsecno \global\subsecno=0
\def\subsec#1{\global\advance\subsecno by1\message{(\secsym\the\subsecno. #1)}
\ifnum\lastpenalty>9000\else\bigbreak\fi
\noindent{\it\secsym\the\subsecno. #1}\writetoca{\string\quad
{\secsym\the\subsecno.} {#1}}\par\nobreak\medskip\nobreak}
\def\appendix#1#2{\global\meqno=1\global\subsecno=0\xdef\secsym{\hbox{#1.}}
\bigbreak\bigskip\noindent{\bf Appendix #1. #2}\message{(#1. #2)}
\writetoca{Appendix {#1.} {#2}}\par\nobreak\medskip\nobreak}
%
%
\def\eqnn#1{\xdef #1{(\secsym\the\meqno)}\writedef{#1\leftbracket#1}%
\global\advance\meqno by1\wrlabeL#1}
\def\eqna#1{\xdef #1##1{\hbox{$(\secsym\the\meqno##1)$}}
\writedef{#1\numbersign1\leftbracket#1{\numbersign1}}%
\global\advance\meqno by1\wrlabeL{#1$\{\}$}}
\def\eqn#1#2{\xdef #1{(\secsym\the\meqno)}\writedef{#1\leftbracket#1}%
\global\advance\meqno by1$$#2\eqno#1\eqlabeL#1$$}
%
\newskip\footskip\footskip14pt plus 1pt minus 1pt 
\def\footnotefont{\ninepoint}\def\f@t#1{\footnotefont #1\@foot}
\def\f@@t{\baselineskip\footskip\bgroup\footnotefont\aftergroup\@foot\let\next}
\setbox\strutbox=\hbox{\vrule height9.5pt depth4.5pt width0pt}
\global\newcount\ftno \global\ftno=0
\def\foot{\global\advance\ftno by1\footnote{$^{\the\ftno}$}}
%
\newwrite\ftfile
\def\footend{\def\foot{\global\advance\ftno by1\chardef\wfile=\ftfile
$^{\the\ftno}$\ifnum\ftno=1\immediate\openout\ftfile=foots.tmp\fi%
\immediate\write\ftfile{\noexpand\smallskip%
\noexpand\item{f\the\ftno:\ }\pctsign}\findarg}%
\def\footatend{\vfill\eject\immediate\closeout\ftfile{\parindent=20pt
\centerline{\bf Footnotes}\nobreak\bigskip\input foots.tmp }}}
\def\footatend{}
%
%
\global\newcount\refno \global\refno=1
\newwrite\rfile
\def\ref{[\the\refno]\nref}
\def\nref#1{\xdef#1{[\the\refno]}\writedef{#1\leftbracket#1}%
\ifnum\refno=1\immediate\openout\rfile=refs.tmp\fi
\global\advance\refno by1\chardef\wfile=\rfile\immediate
\write\rfile{\noexpand\item{#1\
}\reflabeL{#1\hskip.31in}\pctsign}\findarg}
\def\findarg#1#{\begingroup\obeylines\newlinechar=`\^^M\pass@rg}
{\obeylines\gdef\pass@rg#1{\writ@line\relax #1^^M\hbox{}^^M}%
\gdef\writ@line#1^^M{\expandafter\toks0\expandafter{\striprel@x #1}%
\edef\next{\the\toks0}\ifx\next\em@rk\let\next=\endgroup\else\ifx\next\empty%
\else\immediate\write\wfile{\the\toks0}\fi\let\next=\writ@line\fi\next\relax}}
\def\striprel@x#1{} \def\em@rk{\hbox{}}
\def\lref{\begingroup\obeylines\lr@f}
\def\lr@f#1#2{\gdef#1{\ref#1{#2}}\endgroup\unskip}
\def\semi{;\hfil\break}
\def\addref#1{\immediate\write\rfile{\noexpand\item{}#1}} 
\def\startrefs#1{\immediate\openout\rfile=refs.tmp\refno=#1}
\def\xref{\expandafter\xr@f}\def\xr@f[#1]{#1}
\def\refs#1{\count255=1[\r@fs #1{\hbox{}}]}
\def\r@fs#1{\ifx\und@fined#1\message{reflabel \string#1 is undefined.}%
\nref#1{need to supply reference \string#1.}\fi%
\vphantom{\hphantom{#1}}\edef\next{#1}\ifx\next\em@rk\def\next{}%
\else\ifx\next#1\ifodd\count255\relax\xref#1\count255=0\fi%
\else#1\count255=1\fi\let\next=\r@fs\fi\next}
%

%
\newwrite\ffile\global\newcount\figno \global\figno=1
\def\fig{fig.~\the\figno\nfig}
\def\nfig#1{\xdef#1{fig.~\the\figno}%
\writedef{#1\leftbracket fig.\noexpand~\the\figno}%
\ifnum\figno=1\immediate\openout\ffile=figs.tmp\fi\chardef\wfile=\ffile%
\immediate\write\ffile{\noexpand\medskip\noexpand\item{Fig.\
\the\figno. }
\reflabeL{#1\hskip.55in}\pctsign}\global\advance\figno
by1\findarg}
\def\xfig{\expandafter\xf@g}\def\xf@g fig.\penalty\@M\ {}
\def\figs#1{figs.~\f@gs #1{\hbox{}}}
\def\f@gs#1{\edef\next{#1}\ifx\next\em@rk\def\next{}\else
\ifx\next#1\xfig #1\else#1\fi\let\next=\f@gs\fi\next}
\newwrite\lfile
{\escapechar-1\xdef\pctsign{\string\%}\xdef\leftbracket{\string\{}
\xdef\rightbracket{\string\}}\xdef\numbersign{\string\#}}

\def\writestop{\def\writestoppt{\immediate\write\lfile{\string\pageno%
\the\pageno\string\startrefs\leftbracket\the\refno\rightbracket%
\string\def\string\secsym\leftbracket\secsym\rightbracket%
\string\secno\the\secno\string\meqno\the\meqno}\immediate\closeout\lfile}}
\def\writestoppt{}\def\writedef#1{}
\def\seclab#1{\xdef #1{\the\secno}\writedef{#1\leftbracket#1}\wrlabeL{#1=#1}}
\def\subseclab#1{\xdef #1{\secsym\the\subsecno}%
\writedef{#1\leftbracket#1}\wrlabeL{#1=#1}}
\newwrite\tfile \def\writetoca#1{}
\def\leaderfill{\leaders\hbox to 1em{\hss.\hss}\hfill}
\def\writetoc{\immediate\openout\tfile=toc.tmp
     \def\writetoca##1{{\edef\next{\write\tfile{\noindent ##1
     \string\leaderfill {\noexpand\number\pageno} \par}}\next}}}

%
\catcode`\@=12 
%
\edef\tfontsize{\ifx\answ\bigans scaled\magstep3\else
scaled\magstep4\fi} \font\titlerm=cmr10 \tfontsize
\font\titlerms=cmr7 \tfontsize \font\titlermss=cmr5 \tfontsize
\font\titlei=cmmi10 \tfontsize \font\titleis=cmmi7 \tfontsize
\font\titleiss=cmmi5 \tfontsize \font\titlesy=cmsy10 \tfontsize
\font\titlesys=cmsy7 \tfontsize \font\titlesyss=cmsy5 \tfontsize
\font\titleit=cmti10 \tfontsize \skewchar\titlei='177
\skewchar\titleis='177 \skewchar\titleiss='177
\skewchar\titlesy='60 \skewchar\titlesys='60
\skewchar\titlesyss='60
\def\titlefont{\def\rm{\fam0\titlerm}
\textfont0=\titlerm \scriptfont0=\titlerms
\scriptscriptfont0=\titlermss \textfont1=\titlei
\scriptfont1=\titleis \scriptscriptfont1=\titleiss
\textfont2=\titlesy \scriptfont2=\titlesys
\scriptscriptfont2=\titlesyss \textfont\itfam=\titleit
\def\it{\fam\itfam\titleit}\rm}
 \ifx\answ\bigans\else scaled\magstep1\fi
\ifx\answ\bigans\def\abstractfont{\tenpoint}\else
\font\abssl=cmsl10 scaled \magstep1 \font\absrm=cmr10
scaled\magstep1 \font\absrms=cmr7 scaled\magstep1
\font\absrmss=cmr5 scaled\magstep1 \font\absi=cmmi10
scaled\magstep1 \font\absis=cmmi7 scaled\magstep1
\font\absiss=cmmi5 scaled\magstep1 \font\abssy=cmsy10
scaled\magstep1 \font\abssys=cmsy7 scaled\magstep1
\font\abssyss=cmsy5 scaled\magstep1 \font\absbf=cmbx10
scaled\magstep1 \skewchar\absi='177 \skewchar\absis='177
\skewchar\absiss='177 \skewchar\abssy='60 \skewchar\abssys='60
\skewchar\abssyss='60
\def\abstractfont{\def\rm{\fam0\absrm}
\textfont0=\absrm \scriptfont0=\absrms \scriptscriptfont0=\absrmss
\textfont1=\absi \scriptfont1=\absis \scriptscriptfont1=\absiss
\textfont2=\abssy \scriptfont2=\abssys \scriptscriptfont2=\abssyss
\textfont\itfam=\bigit \def\it{\fam\itfam\bigit}\def\footnotefont{\tenpoint}%
\textfont\slfam=\abssl \def\sl{\fam\slfam\abssl}%
\textfont\bffam=\absbf \def\bf{\fam\bffam\absbf}\rm}\fi
\def\tenpoint{\def\rm{\fam0\tenrm}
\textfont0=\tenrm \scriptfont0=\sevenrm \scriptscriptfont0=\fiverm
\textfont1=\teni  \scriptfont1=\seveni  \scriptscriptfont1=\fivei
\textfont2=\tensy \scriptfont2=\sevensy \scriptscriptfont2=\fivesy
\textfont\itfam=\tenit \def\it{\fam\itfam\tenit}\def\footnotefont{\ninepoint}%
\textfont\bffam=\tenbf
\def\bf{\fam\bffam\tenbf}\def\sl{\fam\slfam\tensl}\rm}
\font\ninerm=cmr9 \font\sixrm=cmr6 \font\ninei=cmmi9
\font\sixi=cmmi6 \font\ninesy=cmsy9 \font\sixsy=cmsy6
\font\ninebf=cmbx9 \font\nineit=cmti9 \font\ninesl=cmsl9
\skewchar\ninei='177 \skewchar\sixi='177 \skewchar\ninesy='60
\skewchar\sixsy='60
\def\ninepoint{\def\rm{\fam0\ninerm}
\textfont0=\ninerm \scriptfont0=\sixrm \scriptscriptfont0=\fiverm
\textfont1=\ninei \scriptfont1=\sixi \scriptscriptfont1=\fivei
\textfont2=\ninesy \scriptfont2=\sixsy \scriptscriptfont2=\fivesy
\textfont\itfam=\ninei \def\it{\fam\itfam\nineit}\def\sl{\fam\slfam\ninesl}%
\textfont\bffam=\ninebf \def\bf{\fam\bffam\ninebf}\rm}
%
%

\hyphenation{anom-aly anom-alies coun-ter-term coun-ter-terms}
\def\inv{^{\raise.15ex\hbox{${\scriptscriptstyle -}$}\kern-.05em 1}}

\def\Dsl{\,\raise.15ex\hbox{/}\mkern-13.5mu D} 
\def\dsl{\raise.15ex\hbox{/}\kern-.57em\partial}

 \def\Tr{{\rm Tr}}
\font\bigit=cmti10 scaled \magstep1
\def\lspace{\ifx\answ\bigans{}\else\qquad\fi}
\def\lbspace{\ifx\answ\bigans{}\else\hskip-.2in\fi} 
\def\boxeqn#1{\vcenter{\vbox{\hrule\hbox{\vrule\kern3pt\vbox{\kern3pt
      \hbox{${\displaystyle #1}$}\kern3pt}\kern3pt\vrule}\hrule}}}
\def\mbox#1#2{\vcenter{\hrule \hbox{\vrule height#2in
          \kern#1in \vrule} \hrule}}  
%
 \def\CO{{\cal O}} 
\def\CA{{\cal A}}   
 \def\CH{{\cal H}}  
 \def\CR{{\cal R}} \def\CD{{\cal D}} 
\def\e#1{{\rm e}^{^{\textstyle#1}}}

\def\darr#1{\raise1.5ex\hbox{$\leftrightarrow$}\mkern-16.5mu #1}

\def\half{{\textstyle{1\over2}}} 
\def\roughly#1{\raise.3ex\hbox{$#1$\kern-.75em\lower1ex\hbox{$\sim$}}}

\def\np#1#2#3{Nucl. Phys. {\bf B#1} (#2) #3}
\def\pl#1#2#3{Phys. Lett. {\bf #1B} (#2) #3}

\def\anp#1#2#3{Ann. Phys. {\bf #1} (#2) #3}
\def\pr#1#2#3{Phys. Rev. {\bf #1} (#2) #3}
\def\ap#1#2#3{Ann. Phys. {\bf #1} (#2) #3}

\def\cmp#1#2#3{Comm. Math. Phys. {\bf #1} (#2) #3}
\def\mpl#1#2#3{Mod. Phys. Lett. {\bf #1} (#2) #3}

\def\jhep#1#2#3{JHEP {\bf#1}(#2) #3}

\def\ijmp#1#2#3{Int.~J.~Mod.~Phys. {\bf #1} (#2) #3}
\def\atmp#1#2#3{Adv.~Theor.~Math.~Phys.{\bf #1} (#2) #3}
\def\ap#1#2#3{Ann.~Phys. {\bf #1} (#2) #3}
\def\IB{\relax\hbox{$\inbar\kern-.3em{\rm B}$}}
\def\IC{{\bf C}}
\def\ID{\relax\hbox{$\inbar\kern-.3em{\rm D}$}}
\def\IE{\relax\hbox{$\inbar\kern-.3em{\rm E}$}}
\def\IF{\relax\hbox{$\inbar\kern-.3em{\rm F}$}}
\def\IG{\relax\hbox{$\inbar\kern-.3em{\rm G}$}}
\def\IGa{\relax\hbox{${\rm I}\kern-.18em\Gamma$}}
\def\IH{\relax{\rm I\kern-.18em H}}
\def\IK{\relax{\rm I\kern-.18em K}}
\def\IL{\relax{\rm I\kern-.18em L}}
\def\IP{\relax{\rm I\kern-.18em P}}
\def\IR{{\bf R}}
\def\IZ{{\bf Z}}
\def\II{\relax{\rm I\kern-.18em I}}

\def\ndt{{\noindent}}


\def\CA{{\cal A}}

\def\CD{{\cal D}}

\def\CH{{\cal H}}

\def\CK{{\cal K}}

\def\CN{{\cal N}}
\def\CO{{\cal O}}

\def\CR{{\cal R}}

\def\CZ{{\cal Z}}

\def\p{\partial}
\def\pb{\bar{\partial}}

\def\dir{\rlap{\hskip0.2em/}D}



\def\Tr{{\rm Tr}}

\def\sdtimes{\mathbin{\hbox{\hskip2pt\vrule height 4.1pt depth -.3pt
width.25pt\hskip-2pt$\times$}}}


\def\inbar{\,\vrule height1.5ex width.4pt depth0pt}

\def\sdtimes{\mathbin{\hbox{\hskip2pt\vrule height 4.1pt
depth -.3pt width .25pt\hskip-2pt$\times$}}}
\def\a{{\alpha}}
\def\ap{{\a}^{\prime}}
\def\b{{\beta}}
\def\d{{\delta}}
\def\g{{\gamma}}
\def\e{{\epsilon}}
\def\z{{\zeta}}
\def\ve{{\varepsilon}}
\def\vf{{\varphi}}
\def\m{{\mu}}
\def\n{{\nu}}

\def\l{{\lambda}}
\def\s{{\sigma}}
\def\t{{\theta}}
\def\vt{{\vartheta}}
\def\o{{\omega}}
\def\nc{noncommutative\ }

\def\lref{\begingroup\obeylines\lr@f}
\def\lr@f#1#2{\gdef#1{\ref#1{#2}}\endgroup\unskip}
 \lref\ikkt{N.~Ishibashi, H.~Kawai, Y.~Kitazawa,
and A.~Tsuchiya, \np{498}{1997}{467}, hep-th/9612115}
\lref\curtjuan{C.~G.~Callan, Jr., J.~M.~Maldacena,
\np{513}{1998}{198-212}, hep-th/9708147} \lref\bak{D.~Bak,
\pl{471}{1999}{149-154}, hep-th/9910135} \lref\baklee{D.~Bak,
K.~Lee, hep-th/0007107} \lref\moriyama{S.~Moriyama,
hep-th/0003231} \lref\wadia{A.~Dhar, G.~Mandal and S.~R.~Wadia,
\mpl{A7}{1992}{3129-3146}\semi A.~Dhar, G.~Mandal and S.~R.~Wadia,
\ijmp{A8}{1993}{3811-3828}\semi A.~Dhar, G.~Mandal and
S.~R.~Wadia, \mpl{A8}{1993}{3557-3568}\semi A.~Dhar, G.~Mandal and
S.~R.~Wadia, \pl{329}{1994}{15-26}}

\lref\samson{S.~Shatashvili, hep-th/0105076}

\lref\natibi{N.~Seiberg, hep-th/0008013}

\lref\alexiscs{A.~Polychronakos, hep-th/0007043}

\lref\mateos{D.~Mateos, ``Noncommutative vs. commutative
descriptions of D-brane BIons'', hep-th/0002020}
\lref\mrs{S.~Minwala, M.~ van Raamsdonk, N.~Seiberg,
``Noncommutative Perturbative Dynamics'', hep-th/9912072}
\lref\nahm{W.~Nahm, \pl{90}{1980}{413}\semi W.~Nahm, ``The
Construction of All Self-Dual Multimonopoles by the ADHM Method'',
in ``Monopoles in quantum field theory'', Craigie et al., Eds.,
World Scientific, Singapore (1982) \semi N.J.~Hitchin,
\cmp{89}{1983}{145}} \lref\rs{M.~van Raamsdonk, N.~Seiberg,
``Comments of Noncommutative Perturbative Dynamics'',
hep-th/0002186, \jhep{0003}{2000}{035}} \lref\k{M.~Kontsevich,
``Deformation quantization of Poisson manifolds'', q-alg/9709040}
\lref\gms{R.~Gopakumar, S.~Minwala, A.~Strominger, hep-th/0003160,
\jhep{0005}{2000}{020}} \lref\agms{M.~Aganagic, R.~Gopakumar,
S.~Minwala, A.~Strominger, hep-th/0009142 } \lref\sst{N.~Seiberg,
L.~Susskind, N.~Toumbas, hep-th/0005040} \lref\sdual{R.~Gopakumar,
S.~Minwala, J.~Maldacena, A.~Strominger, hep-th/0005048\semi
O.~Ganor, G.~Rajesh, S.~Sethi, hep-th/00050046}
\lref\filk{T.~Filk, ``Divergencies in a Field Theory on  Quantum
Space'', \pl{376}{1996}{53}} \lref\cf{A.~Cattaneo, G.~Felder, ``A
Path Integral Approach to the Kontsevich Quantization Formula'',
math.QA/9902090}

\lref\cds{A.~Connes, M.~Douglas, A.~Schwarz,
\jhep{9802}{1998}{003}} \lref\douglashull{M.~Douglas, C.~Hull,
 \jhep{9802}{1998}{008},
hep-th/9711165} \lref\wtnc{E.~Witten, \np{268}{1986}{253}}
\lref\volker{V.~Schomerus, \jhep{9906}{1999}{030}}
\lref\cg{E.~Corrigan, P.~Goddard,  \anp {154}{1984}{253}}

\lref\donaldson{S.K.~Donaldson, ``Instantons and Geometric
Invariant Theory", \cmp{93}{1984}{453-460}}

\lref\nakajima{H.~Nakajima, ``Lectures on Hilbert Schemes of
Points on Surfaces''\semi AMS University Lecture Series, 1999,
ISBN 0-8218-1956-9. }

\lref\neksch{N.~Nekrasov, A.~S.~Schwarz, hep-th/9802068,
\cmp{198}{1998}{689}}

\lref\nekrev{N.~Nekrasov, hep-th/0010017} \lref\freck{A.~Losev,
N.~Nekrasov, S.~Shatashvili, ``The Freckled Instantons'', {\tt
hep-th/9908204}, Y.~Golfand Memorial Volume, M.~Shifman Eds.,
World Scientific, Singapore, in press}

\lref\rkh{N.J.~Hitchin, A.~Karlhede, U.~Lindstrom, and M.~Rocek,
\cmp{108}{1987}{535}}

\lref\alexios{A.~Polychronakos, hep-th/0007043}
\lref\branek{H.~Braden, N.~Nekrasov, hep-th/9912019\semi
K.~Furuuchi, hep-th/9912047} \lref\davakisun{D.~Gross,
A.~Hashimoto, N.~Itzhaki, ``Observables in Non-commutative Gauge
Theories'', hep-th/0008075} \lref\wilsloops{N.~Ishibashi, S.~Iso,
H.~Kawai, Y.~Kitazawa, hep-th/9910004, \np{573}{2000}{573-593}
\semi J.~Ambjorn, Y.M.~Makeenko, J.~Nishimura, R.J.~Szabo,
hep-th/9911041, \jhep{9911}{1999}{029} ; hep-th/0002158,
\pl{480}{2000}{399-408}; hep-th/0004147, \jhep{0005}{2000}{023}}

\lref\wilson{G.~ Wilson, ``Collisions of Calogero-Moser particles
and adelic Grassmannian", Invent. Math. 133 (1998) 1-41.}

\lref\polyakov{A.~Polyakov, hep-th/0110196}

\lref\gkp{S.~Gukov, I.~Klebanov, A.~Polyakov, hep-th/9711112,
\pl{423}{1998}{64-70}} \lref\abs{O.~Aharony, M.~Berkooz,
N.~Seiberg, hep-th/9712117, \atmp{2}{1998}{119-153}}

\lref\abkss{O.~Aharony, M.~Berkooz, S.~Kachru, N.~Seiberg,
E.~Silverstein, hep-th/9707079, \atmp{1}{1998}{148-157}}

\lref\witsei{N.~Seiberg, E.~Witten, hep-th/9908142,
\jhep{9909}{1999}{032}} \lref\kinks{E.~Teo, C.~Ting, ``Monopoles,
vortices and kinks in the framework of noncommutative geometry'',
\pr{D56}{1997}{2291-2302}, hep-th/9706101} \lref\kkn{V.~Kazakov,
I.~Kostov, N.~Nekrasov, ``D-particles, Matrix Integrals and KP
hierachy'', \np{557}{1999}{413-442}, hep-th/9810035}
\lref\tdgt{papers on 2d YM} \lref\manuel{D.-E.~Diaconescu,
\np{503}{1997}{220-238}, hep-th/9608163}

\lref\genmnp{L.~Jiang, ``Dirac Monopole in Non-Commutative
Space'', hep-th/0001073} \lref\hashimoto{K.~Hashimoto, H.~Hata,
S.~Moriyama, hep-th/9910196, \jhep{9912}{1999}{021}\semi
A.~Hashimoto, K.~Hashimoto, hep-th/9909202,
\jhep{9911}{1999}{005}\semi K.~Hashimoto, T.~Hirayama,
hep-th/0002090} \lref\hklm{J.~Harvey, P.~Kraus, F.~Larsen,
E.~Martinec, hep-th/0005031}

\lref\snyder{H.~S.~Snyder, ``Quantized Space-Time'',
\pr{71}{1947}{38}; ``The Electromagnetic Field in Quantized
Space-Time'', \pr{72}{1947}{68}} \lref\connes{A.~Connes,
``Noncommutative geometry'', Academic Press (1994)}
\lref\planar{A.~Gonzalez-Arroyo,  C.P.~Korthals Altes, ``Reduced
model for large $N$ continuum field theories'',
\pl{131}{396}{1983}} \lref\barsminic{I.~Bars, D.~Minic,
``Non-Commutative Geometry on a Discrete Periodic Lattice and
Gauge Theory'', hep-th/9910091} \lref\grossneki{D.~Gross,
N.~Nekrasov, \jhep{0007}{2000}{034}, hep-th/0005204}
\lref\grossnekii{D.~Gross, N.~Nekrasov,  hep-th/9907204}
\lref\ooguriokawa{H.~Ooguri, Y.~Okawa, hep-th/0103124}

\lref\review{M.~Douglas, N.~Nekrasov,  hep-th/0106048}

\lref\rcft{A.~Recknagel, V.~Schomerus, hep-th/9712186\semi
G.~Felder, J.~Fr\"olich, J.~Fuchs, C.~Schweigert, hep-th/9909140,
hep-th/9912239}

\lref\dwzw{A.~Alekseev, V.~Schomerus, hep-th/9812193 }

\lref\technique{A.~Schwarz, hep-th/0102182}

\def\ihes{{\it Institute des Hautes Etudes Scientifiques, Le
Bois-Marie, Bures-sur-Yvette, F-91440 France}}
\def\itep{{ Institute for Theoretical and
Experimental Physics, 117259 Moscow, Russia}}

\Title{\vbox{\baselineskip 10pt
\hbox{IHES-P/02/14}\hbox{ITEP-TH-13/02} \hbox{hep-th/0203109}
 {\hbox{
}}}} {\vbox{\vskip -30 true pt \centerline{LECTURES ON OPEN
STRINGS,}
\smallskip\smallskip\centerline{AND NONCOMMUTATIVE GAUGE THEORIES}
\smallskip
\medskip
\vskip4pt }} \vskip -20 true pt \centerline{Nikita
A.~Nekrasov\foot{On leave of absence from \itep}}
\smallskip\smallskip
\centerline{\ihes}
\medskip \centerline{\tt e-mail: nikita@ihes.fr}
\bigskip
\centerline{\it To Alain Connes on his 55th birthday}

\bigskip
The background independent formulation of the gauge theories on
D-branes in flat space-time is considered, some examples of the
solutions of their equations of motion are presented, the
solutions of Dirac equation in these backgrounds are analyzed, and
the generalizations to the curved spaces, like orbifolds,
conifolds, and K3 surfaces, are discussed.

\newsec{Introduction}

We suggest to look for a generalized version of the gauge
fields/strings correspondence \polyakov\ which may prove easier to
establish (as sometimes general problems are easier then the
particular ones) . In the past few years a (new) connection has
been (re-)discovered between noncommutative geometry \connes\ and
string theory \cds\douglashull\volker\witsei. The previous
understanding \wtnc\ of intrinsic noncommutativity of open string
theory was supplemented by a vast number of examples stemming from
the studies of D-branes, which allowed to make the
noncommutativity manifest already at the field theory (or zero
slope) limit. This connection may prove useful both ways. On the
one hand, the noncommutative geometry is a deeply studied subject,
thanks to the work of A.~Connes and his followers. On the other
hand, using the intuition/results from D-brane physics one can
come up with new solutions/ideas for theories on noncommutative
spaces.

These lecture notes should not be considered as an introduction
into the noncommutative field theories and their relation to
string theories. We refer the interested reader to \review.
Instead, we shall expand on several points not covered in \review.

We shall start with a unified construction of the worldvolume
theories of D-branes of unspecified dimensionality. We shall
discuss mostly the flat Minkowski space closed string background.
In the concluding section we present a few new results on curved
backgrounds: namely the orbifolds of flat space, and their
deformations, the conifolds. The more abstract approach to
classification of D-brane states on rational conformal field
theories will not be covered here (see cf. \dwzw\rcft)

In the mean time we shall discuss several classical solutions of
the noncommutative gauge theory. They correspond to flat D-branes,
D-branes at angles, D-branes with different magnetic fields turned
on, D-branes of various dimensions. Generically these solutions
are unstable (open string spectrum contains a tachyon). In the
noncommutative gauge theory the instability is reflected by the
negative modes in the expansion around the solution. In principle
one should be able to study the decay of the solution towards the
stable ones. However the application of this analysis to string
theory is limited, as in the majority of the cases the
${\ap}$-corrections to the flow may not be negligible.

We shall also consider in some detail the case of four dimensional
instanton/monopole solutions. In each case we shall construct the
solutions of the Dirac equation in the background of the
instanton/monopole. In the case of instantons the analysis of the
solutions of the Dirac equation permits to establish the
completeness of the noncommutative version of ADHM construction
\neksch. This is a noncommutative version of the reciprocity of
\cg.

In the course of these lectures we shall try to make clear that
the noncommutative algebras, which should be thought of the
algebras of functions on the noncommutative manifold, are not
fixed, but rather arise upon a choice of classical solution in the
background independent formulation of the gauge theory. We shall
show that this background independent formulation is on the one
hand related to the background independent open string field
theory \samson. On the other hand it is related to Matrix theory.

Finally, a note for Alain. In these notes we look at but the
simplest noncommutative geometries arising in open string theory
on flat space-time backgrounds. Nevertheless, even this simplest
case turns out to be rather rich. In the $\ap \to 0$ limit the
associative algebra which governs the story is the so-called
Yang-Mills algebra. Almost nothing is known about its
representation theory.

\ndt{\bf Acknowledgements.} I would like to thank A.~Connes,
D.~Gross,  A.~Polyakov, A.~Schwarz, S.~Shatashvili for fruitful
discussions. Research was supported in part by RFFI under the
grant 01-01-00549, by Clay Mathematical Institute. I would like to
thank my fellow co-organizers of the 2001 Les Houches summer
school for the opportunity to present these results, and the
students for the useful back-reaction.

\newsec{Background Independence}

It was observed by many authors, see cf. \natibi\alexiscs\ that
the Matrix theory action (or its euclidean version \ikkt) provides
a background independent formulation of the large class of gauge
theories. We shall now state this more precisely, and at the same
time we shall fix our notations.
\def\cy{{\rm g}_{\rm YM}^2}
Consider an abstract Hilbert space ${\CH}$ and a collection of $d$
Hermitian  operators $Y^i$, $i=1, \ldots, d$, acting there. We
shall sometimes use the notation ${\bf Y}$ for the vector in $V =
{\IR}^d$ with values in the space ${\bf H}({\CH})$ of Hermitian
operators in ${\CH}$. Let $g_{ij}$ be a Euclidean metric on $V$,
Define a formal action: \eqn\ma{S = - {1\over{4 {\cy}}} \ \sqrt{g}
\ g_{ik} \ g_{jl} \ {\Tr}_{\CH} [ Y^i, Y^j] [Y^k, Y^l]}We said
\ma\ is formal because it may happen that for a physically
sensible choice of ${\bf Y}$ the value of $S$ is infinite. We
shall consider ${\bf Y}$ such that with appropriate choice of a
constant ${\m}$, the action becomes finite upon subtraction of
${\m} \ {\Tr}_{\CH} {\bf 1}$ (more precisely, we subtract ${\mu}
{\bf 1}$ from $[Y, Y]^2$ before calculating the trace).

The action \ma\ has an obvious gauge symmetry (which is consistent with the
infrared regularization above):
\eqn\gt{Y^i \mapsto g^{\dagger} Y^i g, \ g^{\dagger} g = g g^{\dagger} = {\bf 1}}

The equations of motion following from  \ma\ are: \eqn\eom{g_{kl}
[ Y^k, [Y^l, Y^j]] = 0, \ j = 1, \ldots d}

The operators $Y^i$ generate some associative algebra ${\CA}$. Let
us define the so-called {\it Yang-Mills algebra} ${\CA}_{YM, d}$
to be the associative algebra generated by $Y^i$, subject to the
relations \eom.

We shall now pause to establish the meaning of the operators $Y^i$
in the open string theory context.

To this end let us consider open superstring propagating in the
flat ${\IR}^{9,1}$. Consider the worldsheet of the disk topology.
Let $z$, $\vert z \vert \leq 1$ be the holomorphic coordinate on
this disk $D$. Let $z = e^{-r + i {\s}}$. The fields $X^{\m}$ will
have the following boundary conditions: \eqn\bndryc{\eqalign{&
{\p}_{r} X^{\m} = 0, \qquad {\m} = 0, 1 \ldots, p, \qquad r=1 \cr
& {\p}_{\s} X^{\m} = 0, \qquad {\m} = p+1, \ldots, 25, \qquad r=1
\cr}} These  boundary conditions describe the single flat
$Dp$-brane. If we want to have several parallel $Dp$-branes then
we should allow for the Chan-Paton factors, say $i=1, \ldots, N$,
so that the (constant) value of $X^{\m} (r=1, {\s}) = {\vf}^{\m}$,
${\m} = p+1, \ldots 25$ may depend on $i$: ${\vf}^{\m}_i$.

The canonical way of setting up a string calculation in these
circumstances is to consider the dimension $1$ vertex operators on
the boundary, evaluate their correlation function, and integrate
it over the moduli space of points on the boundary of the disk.
One can also add the closed string vertex operators into the
interior of the disk. These operators should have the dimension
$(1,1)$. The open string vertex operators in general change the
boundary conditions, i.e. the boundary conditions corresponding to
the Chan-Paton index $i$ to  the left of the vertex operator may
be followed by the $j$'th boundary condition to the right of the
operator. This is, of course, reflected by the contribution to the
dimension of the operator of the mass squared of the stretched
string: $m_{ij}^2 = \Vert {\vf}^i - {\vf}^j \Vert^2$.

We would like to have a setup in which there is no need to specify
in advance neither the values of ${\vf}^i$, nor $p$ or $N$. All
this data will be encoded in the properties of the operators
$Y^{\m}$, ${\m} = 0, \ldots, 9$, acting in some auxiliary Hilbert
space ${\CH}$.

Consider the correlation function of a closed string vertex
operator ${\CO}$ inserted at the center $z=0$ of the disk $D$, and
the boundary operator, generalizing the usual supersymmetric
Wilson loop: \eqn\bndryop{ {\CZ} \left[ {\CO} \vert Y \right] =
\langle \ {\CO}  \  {\exp} \left( - {1\over{4\pi\ap}} \int_{D}
g_{ij} \left(  {\p} x^i {\pb} x^j + {\psi}^i {\pb} {\psi}^j +
{\tilde\psi}^i {\p} {\tilde\psi}^j \right) \right) \ {\Tr}_{\CH}
\left(  P\exp\oint_{\p D} i  k_i \left( Y^i - x^i {\bf 1} \right)
+ {\vt}_i {\Psi}^i  +  {\vt}_{i} {\vt}_{j} [ Y^i, Y^j] \right)
\rangle } Here $k_i, {\vt}_i$ are the momenta conjugate to $x^i,
{\Psi}^i = {\psi}^i + {\tilde\psi}^i$: $k_i = g_{ij} {\p}_n x^j,
{\vt}_i = {\psi}^i - {\tilde\psi}^i$. For example, for the
graviton\ooguriokawa: ${\CO}_{h} = h_{ij} (p) : {\psi}^i
{\tilde\psi}^j e^{i p \cdot x} : $ (times the ghosts and
superghosts): $$ {\CZ} \left[ {\CO}_{h} \vert Y \right] = g_{kl}
\int dp e^{- i p \cdot x} h_{ij} (p) \int_{0}^{1} ds \ {\Tr}_{\CH}
\ e^{is  p \cdot Y} [ Y^i, Y^k ] e^{i (1-s) p \cdot Y } [ Y^j,
Y^l] + o ({\ap}) $$ It can be shown that the condition that the
boundary interaction \bndryop\ is consistent with the conformal
invariance of the worldsheet sigma model reads as:
\eqn\ymeoma{g_{ij} [ Y^i, [ Y^j, Y^k] ] = 0 ({\ap})} We can think
of \ymeoma\ as defining a one-parametric family of the associative
algebras, generated by $Y^i$. We shall call them the {\it algebras
of functions on the D-brane worldvolume}.

It should be straightforward to establish a direct relation
between the background independent formulation of the
noncommutative gauge theories via $Y^i$'s, and the background
independent open string field theory\samson.

In the sequel we shall study the case ${\ap}  \to 0$, i.e.
Yang-Mills algebras.

An obvious class of {\it Yang-Mills algebras} is provided by the
Heisenberg-Weyl algebras, where the generators $Y^i$ obey the
stronger condition: \eqn\hwa{[Y^i, Y^j] \in {\it center}
({\CA}_{YM, d})}Clearly, for $d=2$ all Yang-Mills algebras are at
the same time Heisenberg-Weyl.
For generic Heisenberg-Weyl solution let us denote
\eqn\cntr{ Z^{ij} = [ Y^i, Y^j],}
we have $[Z^{ij}, Y^k] = 0$ for any $i,j,k$.
Let us diagonalize $Z^{ij}$:
\eqn\splt{{\CH} = \bigoplus_{A} H_{A}, \qquad Z^{ij}\vert_{H_{A}} = i {\t}^{ij}_A
\ \cdot 1_{H_{A}}}
Then each subspace
$H_{A}$ is an irreducible
representation of the Heisenberg algebra
$[x_A^i, x_A^j] = i {\t}^{ij}_A$.

We now wish to describe the spectrum of fluctuations around the
solution \cntr\splt. To this end we can decompose the fluctuation
$y^i = {\d} Y^i$ as follows: \eqn\dcmps{y^i = \sum_{A,B} y^i_{AB},
\qquad y^i_{AB}: H_B \to H_A} We shall also impose the following
gauge condition: \eqn\ggecnd{g_{ij} [Y^i, y^j] = 0 \leftrightarrow
\sum_{A,B} g_{ij} \left( x^i_A y^j_{AB} - y^i_{AB} x^j_B \right) =
0} The linearized fluctuations are governed by the quadratic
approximation to the action: \eqn\fluc{{\CK} y^j = g_{ik} [x^i,
[x^k, y^j]]  + 2 g_{ik} [y^i, Z^{kj}]} which leads to the
following eigenvalue problem for the spectrum of masses:
\eqn\flucc{\eqalign{ {\o}^2 \ y^j_{AB} =  & \left( {\Delta}_A y^j
+ y^j {\Delta}_B - 2 g_{ik} x^i_A y^j_{AB} x^k_B + 2i y^i_{AB}
\left({T}^{j}_{i,B} - {T}^{j}_{i,A} \right) \right) \cr &
{\Delta}_{A} = g_{ij} x^i_{A} x^j_{A} , \qquad T^j_{i,A} =
{\t}^{kj}_{A} g_{ik} \cr}}

We now proceed with some examples.

\subsec{Dolan-Nappi solutions}

This solution describes two branes which could sit on top of each other, and have different magnetic fields turned on.

Set $d=2$, ${\t}^{ij}_{A,B} = {\t}_{A,B} {\epsilon}^{ij}$, $A,B=1,2, i,j=1,2$,
$g_{ij} = {\d}_{ij}$. The operators $\bf Y$ corresponding to this solution are given by:
\eqn\dnsln{Y^1 + i Y^2 = \pmatrix{\sqrt{2\t_1} a_1 & 0 \cr
0 & \sqrt{2\t_2} a_2 }}
where $a_1, a_2$ are the annihilation operators acting in two
(isomorphic) copies ${\CH}_1, {\CH}_2$ of the Hilbert space ${\CH}$.

To facilitate the analysis
let us map the operators $L_{AB}: H_A \to H_B$ to the vectors in the tensor product: $H_{B} \otimes H_{A}^{*} \approx H \otimes H$:
$$
L_{AB} \mapsto \sum_{n_1, n_2} \langle n_2 \vert L_{AB} \vert n_1 \rangle
\quad \vert n_2, n_1 \rangle
$$
Also, introduce the notation ${\z} = y^1_{AB} + i y^2_{AB}, {\bar\z} = y^1_{AB} - i y^2_{AB} $.
First, assume that ${\t}_{A,B} > 0$ and introduce the creation-annihilation operators:
$$
x_{A,B}^1 + i x_{A,B}^2 = \sqrt{2{\t}_{A,B}} a_{1,2}
$$
and  the number operators $n_{\a} = a_{\a}^{+} a_{\a}$.
Then (upon the identification ${\z} \in H_1 \otimes H_2$)
\eqn\kko{{\CK} \matrix{ {\z} \cr {\bar\z} \cr}
= 2{\t}_A ( n_1 + {\half} ) + 2{\t}_B (n_2 + {\half}) -
2\sqrt{{\t}_A {\t}_B} \left( a_1^{+} a_2^{+} + a_1 a_2 \right) \pm 2 ({\t}_A - {\t}_B) \matrix{ {\z} \cr {\bar\z} \cr}}
This operator is conveniently diagonalized by introduction of the
$SU(1,1)$ generators:
\eqn\suoneone{L_{+} = a_1^{+} a_2^{+}, \ L_{-} = a_1 a_2 , L_{0} = {\half}
\left( n_1 + n_2 + 1\right)} and the operator
$$
M = {\half} \left( n_1 - n_2 \right)
$$
Now, the spectrum of
\eqn\haml{{\CK} = 2 ({\t}_A -
{\t}_B ) ( M \pm 1) + 2 ( {\t}_A + {\t}_B ) L_0 - 2 \sqrt{{\t}_A {\t}_B}
( L_{+} + L_{-} )}
depends on whether ${\t}_A$ equals ${\t}_B$ or not.
If ${\t}_A - {\t}_B \neq 0$ then upon a Bogolyubov $SU(1,1)$ transformation
we can bring ${\CK}$ to the form:
\eqn\hamli{2 ({\t}_A - {\t}_B) ( M \pm 1)  + 2 \vert {\t}_A - {\t}_B \vert
L_0}
whose spectrum is (remember that the spectrum of $L_0$ is given by:
$L_0 = {\half} + \vert M \vert + k, \quad k \in {\IZ}_{+}$):
\eqn\spectr{
2 ({\t}_A - {\t}_B)  ( \pm 1 + M) + \vert {\t}_A - {\t}_B \vert
( 1 + 2 \vert M \vert + 2k)}
which contains a tachyonic mode (for ${\z}$ or ${\bar \z}$ depending on the
sign of ${\t}_A - {\t}_B$).

For ${\t}_A = {\t}_B = {\t}$ the operator
${\CK}$ is manifestly positive definite:
$$
{\CK} = 2{\t} b b^{\dagger}, \qquad b = a_1 - a_2^{+}, \ b = a_1^{+} - a_2
$$
and its spectrum is continuous:
$$
{\CK} = 2{\t} \vert {\kappa} \vert^2
$$
with the eigenvectors:
$$
e^{i( {\kappa} a_1^{+} + {\bar\kappa} a_2^{+}) + a_1^{+}a_{2}^{+}}  \vert 0,0\rangle
$$
which in the ordinary, operator, representation correspond to the plane waves:
$$
e^{i({\kappa} a^{\dagger} + {\bar\kappa}a)}
$$

\subsec{Intersecting branes}

As another interesting example of the solution \hwa\ we shall look
at the case $d=4$: \eqn\ints{\eqalign{{\t}^{12}_A = &{\t}_1
{\d}_{A,1}\cr {\t}^{34}_A = & {\t}_2 {\d}_{A,2}\cr} } This
solution describes two branes, each having two dimensions
transverse to another. The operators $Y^i$ corresponding to this
solution have the following block-diagonal form: \eqn\trns{Y^1 + i
Y^2 = \sqrt{2{\t}_1} \pmatrix{ a_1 & 0 \cr 0 & 0} , \ Y^3 + i Y^4
= \sqrt{2\t_2} \pmatrix{ 0 & 0 \cr 0 & a_2}} where we denote by
$a_1, a_2$ the annihilation operators acting in two copies
${\CH}_1, {\CH}_2$ of the Hilbert space ${\CH}$.

In this case the spectrum of fluctuations contains in the $AB$
sector the discrete modes, corresponding to the strings localized
at the intersection of the branes, and, if $\t_1 \neq \t_2$, starts off with
the tachyonic mode:
\eqn\spect{{\o}^2 = 2{\t}_1 n_1 + 2{\t}_2 n_2  \pm \vert {\t}_1 - {\t}_2
\vert , \ n_{1,2} \in {\IZ}_+}

\subsec{T-duality}

The configuration from the previous example is closely related to the `piercing string' solution of \grossnekii, which in our present notation
is described as follows: let as before ${\CH}_1, {\CH}_2$ denote two copies of the Hilbert space ${\CH}$. Now, let ${\CH}_3 = {\CH}_1 \otimes {\CH}_2$.
Then the operators $\bf Y$ will be acting in ${\CH}_3 \oplus {\CH}_2$:
\eqn\prcstr{Y^1 + i Y^2 = \sqrt{2\t_1}
\pmatrix{ a_1 \otimes 1 & 0 \cr 0 & 0},
\ Y^3 + i Y^4 = \sqrt{2\t_2} \pmatrix{ 1 \otimes {\half}( a_2 + a_2^{+}) & 0 \cr
0 & a_2 }  }

The analysis of fluctuations around this solution is similar, it also exibiths
a tachyonic mode for ${\t}_1 \neq {\t}_2$.

\newsec{BPS algebras}

Less trivial is the case of the so-called {\it BPS algebras} (not
to be confused with the algebras of BPS states), whose generators
obey the following relations: let $S_{\pm}$ be the spaces of
chiral spinors of $SO(d)$, let ${\g}_i$ be the generators of the
Clifford algebra, \eqn\cliff{ \{ {\g}_i, {\g}_j \} = g_{ij} \ {\bf
1} , }then the relations state that there exist two spinors
${\e}_1, {\e}_2 \in S_{+}$ such that \eqn\bps{ [Y^i, Y^j] [
{\g}_i, {\g}_j ] \ {\e}_1 + {\bf 1} \ {\e}_2 = 0} If $d=4$ then
these relations have the following simple form: \eqn\asd{ [Y^i,
Y^j] \pm {\half} {\ve}_{mnkl} \sqrt{g} \ g^{mi} \ g^{nj} [ Y^k,
Y^l] \in center ({\CA}_{YM, 4})} The operators $Y^i$ solving \asd,
considered up to a gauge transformation \gt, define the {\it
noncommutative instanton}.

\subsec{Noncommutative U(1) instantons}

The construction of noncommutative instantons \neksch\ is a
generalization of the famous ADHM procedure, which produces an
anti-self-dual gauge field on ${\IR}^4$ (or its one-point
compactification, ${\bf S}^4$, given a solution to a
finite-dimensional version of the anti-self-duality condition. The
latter is imposed on the set of complex matrices, $B_{\a}$, ${\a}
= 1,2$, and (in the $U(1)$ case) $I$, where $B_{\a}$ are the
operators in a vector space $V = {\IC}^k$, where $k$ is the
instanton charge, while $I$ is a vector in $V$. The conditions,
imposed on $(B_{\a}, I)$ are: \eqn\adhme{\eqalign{& [B_{1}, B_{2}
]= 0 \cr & [B_1 , B_1^{\dagger}] + [ B_2, B_2^{\dagger}] +
II^{\dagger} = 2\cdot {\bf 1}_V \cr}} The $2$ in the right-hand
side of \adhme\ is a convenient choice of normalization, which
could be altered -- what matters is whether there stands $0$ or
something positive. Given a solution to \adhme, one can generate
another one by applying a $U(k)$ transformation $(B_{\a}, I)
\mapsto (g^{-1} B_{\a} g, g^{-1} I)$ with $g \in U(k)$. Such
solutions will be considered as equivalent ones.

The purpose of this note is to elucidate the meaning of the space
$V$ from the point of view of four dimensional \nc\ gauge theory.
We shall see that $V$ is nothing but the space of normalizable
solutions to Dirac equation for the spinor field in the
fundamental representation, in the instanton background.

We now recall the construction of the instanton gauge field.
Consider the associative algebra ${\IR}^4_{\t}$ of operators in
the Hilbert space ${\CH}= L^2 ({\IR}^2)$, which we identify with
the space of states of a two-dimensional harmonic oscillator. Let
$a_{\a}$, $a_{\a}^{\dagger}$, ${\a} = 1,2$ be the annihilation and
creation operators, respectively, which obey the algebra:
\eqn\ncr{[a_{\a}, a_{\b}^{\dagger}] = {\d}_{\a\b}} Consider the
${\IR}^4_{\t}$-module $M = {\CH} \otimes V$, and let
\def\D{{\Delta}}
\def\Dt{{\widetilde{\D}}}
$\D$, $\Dt$ denote the operators in $M$:\eqn\Lapl{\eqalign{& {\D}
= \sum_{\a} (B_{\a} - a_{\a}^{\dagger})(B_{\a}^{\dagger} - a_{\a})
\cr & {\Dt} = \sum_{\a} (B_{\a}^{\dagger} - a_{\a}) (B_{\a} -
a_{\a}^{\dagger}) \cr & {\Dt} - {\D} = I I^{\dagger} \cr}} Let
${\CD}^{\dagger} : M \otimes {\IC}^2 \oplus {\CH} \to M \otimes
{\IC}^2$ be a morphism of ${\IR}^4_{\t}$ modules, given by:
\eqn\mor{{\CD}^{\dagger} = \pmatrix{B_1 - a_1^{\dagger} & B_2 -
a_2^{\dagger} & I \cr -B_2^{\dagger} + a_2 & B_1^{\dagger} - a_1 &
0 \cr}}It follows from \adhme\ that ${\CD}^{\dagger} {\CD} = {\Dt}
\otimes {\bf 1}_{{\IC}^2}: M {\otimes} {\IC}^2 \to M {\otimes}
{\IC}^2$. One shows \nekrev\ that ${\Dt}$ is a positive definite
Hermitian operator in $M$. Hence, the following operator is a
well-defined projector in $M {\otimes} {\IC}^2 {\oplus} {\CH}$:
\eqn\prone{{\Pi}_1 = {\CD} {1\over{{\CD}^{\dagger}{\CD}}}
{\CD}^{\dagger}} Let ${\Psi}$ denote the fundamental solution to
the equation ${\CD}^{\dagger} {\Psi} = 0$, i.e. a morphism of
${\IR}^4_{\t}$ modules: ${\Psi} : {\CH} \to M {\otimes} {\IC}^2
{\oplus} {\CH}$. One shows that ${\Psi}$ can be normalized so as
to define a unitary isomorphism between ${\CH}$ and the kernel of
${\CD}^{\dagger}$: ${\Psi}^{\dagger} {\Psi} = {\bf 1}_{\CH}$:
\eqn\ps{\eqalign{& {\Psi} = \pmatrix{{\psi}_{1} \cr {\psi}_2 \cr
{\xi} \cr}, \cr  & {\psi}_{\a}  = (B_{\a}^{\dagger} - a_{\a})v \cr
& {\D} v  = \quad - I {\xi} \cr & {\xi}  =
{\Lambda}^{-\half}S^{\dagger}  \cr & {\Lambda} = 1 + I^{\dagger}
{1\over\D} I = {{I^{\dagger}{1\over\D} I}\over{I^{\dagger}
{1\over\Dt} I}} \cr & {\Lambda}^{-1} = 1 - I^{\dagger} {\Dt}^{-1}
I \cr &  S^{\dagger} : {\CH} \to {\CH}, \quad SS^{\dagger} = 1,
S^{\dagger}S = 1 - P \cr}} where $P$ is a orthogonal projection in
${\CH}$ onto a subspace, isomorphic to $V$, which is spanned by
the elements ${\eta}$ which are in the image of the operator
$I^{\dagger} {\exp} \left( \sum_{\a} B_{\a}^{\dagger}
a_{\a}^{\dagger} \right) \vert 0,0 \rangle \ : V \to {\CH}$, where
$\vert 0,0\rangle$ is the vacuum state in ${\CH}$.

It is instructive to show that this image can be also
characterized as the kernel of the operator ${\Lambda}^{-1}$.
Indeed, let ${\l} \in {\CH}$ be such that ${\l} = I^{\dagger}
{\Dt}^{-1} I {\l}$. It follows:\eqn\lemma{\eqalign{I {\l} =
 I I^{\dagger} {\Dt}^{-1} I {\l} \ & \Rightarrow \ {\D} {\Dt}^{-1}
 I {\l} = \left( {\Dt} - II^{\dagger} \right) {\Dt}^{-1} I {\l}  = 0 \cr
 \ & \Rightarrow {\Dt}^{-1} I {\l} = \exp \left( {\sum_{\a}
 B_\a^{\dagger}a_{\a}^{\dagger} } \right) \vert 0,0\rangle \otimes {\n} , \qquad
 {\n} \in V \cr
 \ & \Rightarrow I \left( {\l} - I^{\dagger}  \exp
 \left(  \sum_{\a} B_{\a}^{\dagger} a_{\a}^{\dagger} \right) \vert 0,0\rangle \otimes {\n} \right) = 0 \cr}}

In writing the formulae \ps\ we only had to use the operator
${\Lambda}^{-1}$ which is an element of ${\IR}^4_{\t}$. However,
for computational purposes it is useful to work with ${\Lambda}$
as well. Technically one has to localize ${\IR}^4_{\t}$ over
${\Lambda}$, i.e. consider the formal polynomials of the form
$\sum_{n} a_n {\Lambda}^n$ where $a_n$ are the elements of
${\IR}^4_{\t}$.

 One defines the second projector: \eqn\prtwo{{\Pi}_2 = {\Psi}
{\Psi}^{\dagger}} It is clear from the positivity of ${\Dt}$,
that: \eqn\prj{{\Pi}_1 + {\Pi}_2 = {\bf 1}_{M {\otimes} {\IC}^2
{\oplus} {\CH}}}This relation implies that the following
identities hold: \eqn\tozhd{\eqalign{& (B_1^{\dagger} - a_1)
{1\over{\D}} (B_1 - a_1^{\dagger}) + (B_2 -
a_2^{\dagger}){1\over{\Dt}} (B_2^{\dagger} - a_2) = {\bf 1}_{k}
\cr & (B_2^{\dagger} - a_2) {1\over{\D}} (B_2 - a_2^{\dagger}) +
(B_1 - a_1^{\dagger}){1\over{\Dt}} (B_1^{\dagger} - a_1) = {\bf
1}_{k}\cr & \left( I^{\dagger} {1\over\Dt}  - {\Lambda}^{-1}
I^{\dagger} {1\over\D} \right) (B_{\a} - a_{\a}^{\dagger}) =
0\cr}}Now, define operators in ${\CH}$ \eqn\ops{A_{\a} =
{\Psi}^{\dagger} a_{\a} {\Psi}, \qquad A_{\a}^{\dagger} =
{\Psi}^{\dagger} a_{\a}^{\dagger} {\Psi}}One shows \neksch\nekrev\
that \eqn\asd{\eqalign{& [A_1, A_2] = 0, \quad [A_1^{\dagger},
A_2^{\dagger}] = 0 \cr & [A_1, A_1^{\dagger}] + [ A_2,
A_2^{\dagger}] = 2 \cr}}

\subsec{Higher dimensional instantons}

We now proceed with discussing BPS solutions involving more then
four $Y$'s. In general, for the $U(N)$ gauge theory on a $p$
complex dimensional K\"ahler manifold $X$ the natural analogues of
the instanton equations are the so-called Hermitian Yang-Mills
equations, which state that the curvature of the gauge field $A$
is of the type $(1,1)$ and that its nonabelian part is primitive,
that is orthogonal to the K\"ahler form ${\o}$:
\eqn\hym{\eqalign{& F^{(2,0)} = 0 \cr & F \wedge {\o}^{p-1} = {\l}
\ {\bf 1} \ {\o}^{p} \cr}} where ${\l}$ is a constant, which can
be computed from the first Chern class of the gauge bundle.

We shall now consider the noncommutative analogues of the
equations \hym. Let as introduce a complex structure on ${\IR}^d$
such that the noncommutativity tensor ${\t}$ is of the type
$(1,1)$. We shall for simplicity assume that it actually related
to the K\"ahler form: ${\t} = {\o}^{-1}$. The equations \hym\ will
now read: \eqn\hymnc{\eqalign{& [Y^{\a}, Y^{\b}] = 0, \ {\a}, {\b}
= 1, \ldots, p \cr & \sum_{\a} \ [ Y^{\a} , Y^{\a ,\dagger} ] = p
\cdot {\bf 1} \cr}} (we have normalized things in such a way that
$Y^{\a} = a_{\a}$, with $[a_{\a}, a_{\b}^{\dagger}] = {\d}_{\a\b}$
is a solution to \hymnc) Then it is relatively easy to produce a
$U(p)$ invariant solution to \hymnc with positive action:
\eqn\newsol{\eqalign{& Y^{\a} = S a_{\a} \left( 1 -
{{p!}\over{(N)_{p}}}\right)^{1\over 2} S^{\dagger} \cr & N =
\sum_{\a} a_{\a, \dagger} a_{\a} \cr & (N)_p = N (N+1) \ldots
(N+p-1) \cr & SS^{\dagger} = {\bf 1}, \ S^{\dagger} S = {\bf 1} -
\vert 0 \rangle \langle 0 \vert \cr}} The action on this solution
is finite: \eqn\act{S_{p} = {\Tr}_{\CH} \left( [ Y^{\a}, Y^{\b,
\dagger}] - {\d}^{\a\b} \right) \left( [Y^{\b}, Y^{\a, \dagger} ]
- {\d}^{\a\b} \right) = p(p-1)}

The topological charges associated with this solution are:

\eqn\topch{\eqalign{& ch_r = {\Tr}_{\CH} F^{\wedge r} \wedge
{\o}^{p - r} \cr & F_{ij}  = {\o}_{ik} {\o}_{jl} \left( [Y^k, Y^l]
- i {\t}^{kl} \right) \cr & ch_1 = 0 \cr & ch_2 = S_{p} = p(p-1)
\cr}}

\newsec{Fermions in the $\bf Y$ backgrounds}

Given a generic ${\bf Y}$ we define {\it Dirac operator} ${\dir} :
S_{+} \otimes {\CA} \longrightarrow S_{-} \otimes {\CA}$ by the
formula: \eqn\dirc{{\dir} {\psi} = {\g}_i [ Y^i,  {\psi}]} Now
consider a background ${\bf Y} = {\underline Y} \oplus {\underline
Y}^{\prime}$, which splits as a direct sum of two independent
$Y$-backgrounds. The Dirac operator, as defined by \dirc\ splits
as a sum of four independent operators. The most interesting for
us is the ``off-diagonal'' one: \eqn\dircf{{\dir} {\chi} = {\g}_i
\left( {\underline Y}^i {\chi} - {\chi} {\underline Y}^{\prime, i}
\right)} Now suppose that ${\underline Y}^{\prime}$ is a frozen
vacuum solution, while ${\underline Y}$ is a dynamical gauge
field. The off-diagonal component $\chi$ of the fermion $\psi$
which enters \dircf\ is what is sometimes called the fermion in
the {\it fundamental} representation (as opposed to the {\it
adjoint} fermion in \dirc).

\subsec{Fermions in the instanton bacgkround}

We are now interested in solution of the Dirac equation in the
instanton background for the fermions in the fundamental
representation: \eqn\drc{{\dir} {\chi} = \pmatrix{{\nabla}_1 & -
{\nabla}_2 \cr {\nabla}_{\bar 1} & {\nabla}_{\bar 2} \cr}
\pmatrix{{\chi}_1 \cr {\chi}_2 \cr} = 0} where the covariant
derivative of a field ${\chi}$ in the (right) fundamental
representation is given by: \eqn\covd{{\nabla}_{\a} {\chi} = -
A_{\a}^{\dagger} {\chi} +{\chi} a_{\a}^{\dagger}, \quad
{\nabla}_{\bar\a} {\chi} =  A_{\a} {\chi} - {\chi} a_{\a}}We claim
that the fundamental solution ${\chi}$ to \drc\ (which should be
thought as of the morphism of the ${\IR}^4_{\t}$ modules: $${\chi}
: M \to {\CH} {\otimes} {\IC}^2 $$upon identification of the space
of solutions with $V$, which is done essentially in the same way
as in \cg) is given by: \eqn\spin{{\chi}_{\a} = v^{\dagger}
(B_{\a} - a_{\a}^{\dagger}) {\Dt}^{-1} = - S {\Lambda}^{\half}
{\p}_{\a} \left( I^{\dagger} {\Dt}^{-1} \right)}It is easy to
check \drc\ using the identities \tozhd. Notice the following
normalization of the solution \spin: \eqn\nrm{\sum_{\a}
{\chi}_{\a}^{\dagger} {\chi}_{\a} = \sum_{\a} [ a_{\a}^{\dagger},
[ a_{\a}, {\Dt}^{-1} ]] \Rightarrow {\Tr}_{\CH} \sum_{\a}
{\chi}^{\dagger}_{\a} {\chi}_{\a} = {\bf 1}_{V}}(in the last
equality we used the fact that the trace of the double commutator
will not change if we set $B_{\a} = B_{\a}^{\dagger} = 0$) By
similar calculations one shows, that: \eqn\recn{{\Tr}_{\CH}
\sum_{\a} {\chi}^{\dagger}_{\a} {\chi}_{\a} \pmatrix{ a_{\b} \cr
a^{\dagger}_{\b} } = \pmatrix{ B_{\b}^{\dagger} \cr B_{\b} }} Now,
given an instanton gauge field $Y^i = (A_{\a}, A_{\a}^{\dagger})$
we consider the associated Dirac operator $\dir$ and the space $V$
of its normalizable zero-modes. We shall require that $Y^i$ obey
the following asymptotics: \eqn\asmpt{Y^i = S x^i S^{\dagger} +
y^i} where $SS^{\dagger} = {\bf 1}$, and the eigenvalues
${\l}_n^i$ of $y^i$ decay faster then ${1\over{n}}$ (after some
reshuffling). Similarly, we are interested in the solutions
${\chi}$ to the Dirac equation, which have the form:
\eqn\asmpti{{\chi}_{\a} = S a_{\a}^{\dagger}
{1\over{(aa^{\dagger})^2}} + \ldots} where $\ldots$ denote
subleading (in the same sense as for the gauge field, except that
with ${1\over{n^2}}$ instead of ${1\over n}$ asymptotics) terms.
One shows, analogously to \cg\ that the space of these solutions
is of the dimension $k$, given by the instanton charge, and that
the formulae \recn\ produce ADHM matrices $B, B^{\dagger}$, while
the asymptotics \eqn\asmtoo{{\chi}_{\a} = - I S a_{\a}^{\dagger}
{1\over{(aa^{\dagger})^2}}} gives $I$. To define ``asymptotics''
of the operators properly, it  is useful to employ the frozen
vacuum solution $\underline{Y}^{\prime}$ and the technique of the
symbols, defined with respect to this solution\foot{Given a
background  $\underline{Y}^{\prime}$ one can define the
generalization of the Weyl symbol of the operator ${\CO}$ by
$f_{\CO} (q) = \int dp \ {\Tr}_{\CH} \left( {\CO} \ {\exp} - i p_i
\left( q^i - Y^{i, \prime} \right) \right)$ and conversely,
${\CO}_{f} = \int \ dp \ dq \ f (q) \ {\exp} \left( i p_i \left(
q^i - Y^{i, \prime} \right) \right)$. However, in general it is
not easy to characterize the class of operators for which these
formulae make sense, and also ${\CO}_{f_{\CO}} \neq {\CO}$},
developed in \technique. By the completely analogous calculations
to \cg\ one proves that $B, B^{\dagger}, I$ obey \adhme. This
establishes the reciprocity.

\subsec{Dirac field in the monopole background}

We now present the solution $\chi = \pmatrix{\chi_+ \cr \chi_{-}}$
to the Dirac equation \eqn\dirone{2 \, D_{c} {\chi}_{+} - (D_3 +
{\Phi} - z) {\chi}_{-} = 0} \eqn\dirtwo{2 \, D_{c}^{\dagger}
{\chi}_{-} - (D_3 - {\Phi} + z) {\chi}_{+} = 0} in the NC monopole
background: \eqn\cgansatz{{\chi}_{\pm} = {\pm}
{\Psi}_{\mp}^{\dagger} {1\over{\Delta}}} where
\eqn\bcp{\eqalign{&\Delta = bb^{\dagger} + c^{\dagger}c\cr &
\matrix{\, \, \, b^{\dagger} \Psi_{+} + c \Psi_{-} = 0\cr
-c^{\dagger} \Psi_{+} + b  {\Psi}_{-} = 0 } \cr}} and the
notations are from \grossneki. First of all, we need to invert
${\Delta}$. This is easy, for \bcp\ implies: \eqn\bcpp{\eqalign{0
= \left( bb^{\dagger} + cc^{\dagger} \right) \Psi_{+} = (\Delta +
1) \Psi_{+} & \quad \Rightarrow \quad {\Delta} {\Psi}_{+} = -
\Psi_{+} \cr 0 = \left( b^{\dagger} b + c^{\dagger}c \right)
\Psi_{-} = ({\Delta} - 1) {\Psi}_{-} & \quad \Rightarrow \quad
{\Delta} {\Psi}_{-} = \Psi_{-} \cr}} However, it is early to
assume that ${\chi}_{\pm} = \Psi_{\mp}^{\dagger}$ since  $\Delta$
has a kernel: $$ {\Delta} f = 0 \Leftrightarrow f = v K$$ where
$v$ is from \grossneki, $v = \sum_{n=0}^{\infty} {\n}_n \, b^{n}
{\vf} \, \, \vert n \rangle\langle n \vert$, and $K$ is an
arbitrary $x_3$-dependent operator in the Fock space ${\CH}$.

Thus, ${\chi}_{\pm} = \Psi_{\mp}^{\dagger} - K^{\dagger}_{\pm}
v^{\dagger}$, and the operators $K$ must be chosen in such a way,
that $\chi_{\pm} \left( z = 0 \right) = 0$, which implies:
\eqn\kop{K_{\pm}^{\dagger} = \Psi_{\mp}^{\dagger}(z=0) {\xi}}

Recall that $\Psi_{+} = c v, \quad \Psi_{-} = - b^{\dagger} v$.
Hence\foot{{\it Proof}: use the identities: ${\xi}v  =
{{b^{n}{\vf}}\over{{\z}_n}},
 {\half}{\p}_3 \left({\xi}v\right)  =  \left( {\p}_z + x_3
- {{{\z}_{n+1}}\over{{\z}_n}} \right) ( {\xi}v) , {\z}_{n+1} =
2x_3 {\z}_n + n {\z}_{n-1}$. Then: ${\chi}_{+} = \left( {\p}_z -
x_3 - {\half} z - {\xi}_n^2 \right) v =  {\half} {\xi}^{-1} \left(
{\p}_3 - z \right) ({\xi}v) = {\half} \left( {\p}_3 + {\Phi} - z
\right) v , {\chi}_{-} = v c^{\dagger} -
{\xi}^{-1}c^{\dagger}{\xi}v =  {\xi}^{-1} {\p}_{c} \left( {\xi} v
\right) = \left( {\p}_c + A_c \right) v$ From this  \dirone\
follows, with the help of the Bogomolny equations obeyed by $A_c,
\Phi$ in \grossneki. It remains to check \dirtwo. Indeed, if we
denote ${\eta} = {\xi}^2, \, {\l}_n (z) = {1\over{{\z}_n}} b^{n}
{\vf}$, then: $\left( {\p}_3 - z \right) {\l}_n = 2{\eta}_{n}
\left( {\l}_{n-1} - {\l}_n \right), \quad {\eta}_n \left( {\p}_3 +
z \right) {1\over{{\eta}_n}} \left( {\p}_3 - z \right) {\l}_n =
4{\eta}_n \left[ {\eta}_{n-1} \left( {\l}_{n-2} - {\l}_{n-1}
\right) + ( z - {\eta}_{n} ) \left( {\l}_{n-1} - {\l}_{n} \right)
\right]$ and at the same time: ${1\over 4} {\eta}
{\p}_{c^{\dagger}} \left( {\eta}^{-1} \left({\p}_{c} {\l}
\right)\right) = {{\eta_n}\over{{\eta}_{n+1}}} (n+1) \left(
{\l}_{n+1} - {\l}_n \right) - n \left( {\l}_n - {\l}_{n-1}
\right)$, which implies \dirtwo}, \eqn\kopp{K_{+}^{\dagger} =
{\xi}^{-1} c^{\dagger} {\xi}, \quad K_{-}^{\dagger} = {\xi}^2}

\newsec{Non-trivial backgrounds}

So far our discussion concerned the flat closed string background.
We shall now generalize the discussion to cover some non-trivial string backgrounds.

\subsec{Orbifolds}

The obvious starting point is to consider orbifolds. So, let
${\Gamma}$ be a discrete subgroup of $Spin(d) \times {\IR}^d$ -
the group of isometries of ${\IR}^d$. For $g \in \Gamma$ let
${\gamma}(g) \in Spin(d)$ be the corresponding rotation matrix,
and $l(g) \in {\IR}^d$ the corresponding shift vector. We have the
following composition rule: \eqn\cmps{( l(g_1) , {\gamma}(g_1))
\times (l(g_2), {\gamma} (g_2)) = (l(g_1) + {\gamma}(g_1) l(g_2),
{\gamma} (g_1) {\gamma}(g_2)) = (l (g_1 g_2), {\gamma} (g_1 g_2))
. } We wish to consider D-branes living in the background obtained
by taking the quotient of the Minkowskian space ${\IR}^d$ by
${\Gamma}$.

The prescription for modifying the action \ma\ to reflect the
orbifolded nature of the ambient space-time is the following. We
demand that the Hilbert space ${\CH}$ forms a representation of
${\Gamma}$. Let ${\Omega}: \Gamma \to {\rm End} ({\CH})$ be the
corresponding homomorphism. If ${\CR}_i$, $i=0, \ldots, r$,  are
irreducible unitary representations of ${\Gamma}$, ${\CR}_0 =
{\IC}$ being the trivial representation, then $$ {\CH} =
\bigoplus_{i} {\CH}_i \otimes {\CR}_i $$ is the decomposition of
${\CH}$. We demand that the operators ${\bf Y}$ are equivariant
with respect to ${\Gamma}$ in the sense that: \eqn\equiv{
{\Omega}^{-1} (g) Y^i {\Omega} (g) = {\gamma}_{j}^{i}(g) Y^j + l^i
(g)} In practice it is convenient to introduce the quiver diagram.
We shall skip some standard issues like the reduced gauge
invariance, and as a consequence more freedom in the action,
coming from the twisted sector fields couplings. The latters are
enumerated by the tensors \eqn\twis{{\t}^{ij}(g) =
{\gamma}^{i}_{i^{\prime}} (h) {\gamma}^{j}_{j^{\prime}} (h)
{\t}^{i^{\prime} j^{\prime}} (h^{-1} g h)} which will alter the
ground state solutions as follows: \eqn\gndst{[ Y^i, Y^j ] = i
{1\over{{\# \Gamma}}}\sum_{g \in {\Gamma}} {\t}^{ij} (g) {\Omega}
(g)} (for infinite ${\Gamma}$ the normalization factor should be
discussed separately) \subsec{Example: ${\Gamma} = {\IZ}_2$} Let
${\Gamma} = {\IZ}_2$ act on ${\IR}^d$ as $y \mapsto - y$. We have:
\eqn\hlb{{\CH} = {\CH}_{+} \oplus {\CH}_{-}} The solution to
\equiv\ reads: \eqn\zitwo{Y^i = \pmatrix{ 0 & Y^i_{+-} \cr
Y^{i}_{-+} & 0 \cr}} where $Y^i_{\pm\mp} : {\CH}_{\mp} \to
{\CH}_{\pm}$. To simplify our problem let us assume very special
form of the twisted sector field: $$ {\t}^{ij} (-1) = {\zeta}
{\t}^{ij} (+1) $$ Then, by going to the complex notations we can
rewrite \gndst\ as follows: \eqn\gndste{[ A_{\a},
A_{\b}^{\dagger}] = {\half} {\d}_{\a\b} (1 + {\z} F), \qquad
[A_\a, A_\b] = 0, {\a}, {\b} = 1, \ldots, d/2} where $F$ is the
parity operator: $F \vert_{{\CH}_{\pm}} = {\pm} 1$.

We shall now study the representation theory of the algebra \gndst. Clearly,
it is sufficient to study the case $d=2$.
Let us rewrite \zitwo\ in terms of $A, A^{\dagger}$:
\eqn\zitwoe{A_{\a} = \pmatrix{0 & b_{\a} \cr a_{\a} & 0 \cr},}
then \gndste\ becomes the condition (we now drop the index $\a$):
\eqn\gndstee{ aa^{\dagger} - b^{\dagger} b = {\half} ( 1- {\z}),
\quad bb^{\dagger} - a^{\dagger} a = {\half} ( 1 + {\z})}
We can assume ${\z} \geq 0$ (otherwise we exchange $a$ and $b$).
Notice that if ${\z} = 0$ we can take as a solution
$$
b = {1\over{{\sqrt{2}}}} (1-P) {\bf a} P, a = {1\over{{\sqrt{2}}}}
P {\bf a} (1-P) , \qquad P = {\half} ( 1 + F), \quad F = (-1)^{{\bf a}^{\dagger} {\bf a}}
$$
and ${\bf a}, {\bf a}^{\dagger}$ are the standard creation-annihilation
operators.

For $0 \leq {\z} < 1$ we have the following (essentially unique) representation:
\eqn\smallz{\eqalign{b e^{+}_{n} = \sqrt{n} e^{-}_{n-1} , \quad & \quad a^{\dagger} e_{n}^{+} = \sqrt{ n + {\half} ( 1- {\z})} e_{n}^{-} \cr
b^{\dagger} e_{n}^{-} = \sqrt{n+1} e_{n+1}^{+} , \quad & \quad a e^{-}_{n} =
\sqrt{n + {\half} ( 1- {\z})} e_{n}^{+} \cr}}
For $ {\z} \geq 1$
\eqn\largez{\eqalign{ b e_{n}^{+} = \sqrt{ n  + {\half} ( {\z} - 1)} e^{-}_{n-1}, \quad & \quad a^{\dagger} e_{n}^{+} = \sqrt{ n } e^{-}_{n},
\cr
 b^{\dagger}
e_{n}^{-} = \sqrt{n  + {\half} ( {\z} + 1)} e^{+}_{n+1} \quad & \quad
a e^{-}_{n} = \sqrt{n} e^{+}_{n},
\cr}}
and the vector $e_{0}^{+}$ should be dropped from the representation.

\subsec{Example: instantons on K3}

The closed string background corresponding to the K3 surface can
be realized as a marginal deformation of the orbifold ${\bf
T}^4/{\IZ}_2$ where ${\IZ}_2$ acts by reflection on all four flat
coordinates on the four-torus. In turn, the four-torus is a
quotient of the Euclidean space ${\IR}^4$ by a lattice
${\Gamma}_{0} \approx {\IZ}_4$. Thus we may hope to realize the
noncommutative gauge theory on K3 by deforming the orbifold of
${\IR}^4$ with respect to ${\Gamma} = {\Gamma}_{0} \sdtimes
{\IZ}_2$. Thus we arrive at the following algebra (here $B$ is an
anti-symmetric form on ${\Gamma}_0$ corresponding to the
$B$-field): \eqn\kthree{\eqalign{& U_{l} U_{l^{\prime}} = U_{l +
l^{\prime}} e^{i {\pi} B (l, l^{\prime})}, \qquad l, l^{\prime}
\in {\Gamma}_0 \cr & U_{-l} Y^i U_{l} = Y^i + l^i \cr &
{\Omega}^{-1} Y^i {\Omega} = - Y^i \cr & {\Omega}^2 = 1, \qquad
{\Omega}^{-1} U_{l} {\Omega} = U_{-l} \cr}}and the vacuum
equations \eqn\kthreevc{[Y^i , Y^j ] = {\m}^{ij} + \sum_{l \in
{\Gamma}_0} {\m}^{ij}_l U_l {\Omega}}For consistency the
parameters ${\m}^{ij}_l$ must obey: $\sum_l {\m}^{ij}_l U_l
{\Omega} = \sum_l {\m}^{ij}_l U_{-l^{\prime}} U_{l} {\Omega}
U_{l^{\prime}} = \sum_l {\m}^{ij}_l U_{l - 2l^{\prime}} {\Omega}
$, i.e. ${\m}^{ij}_l = {\m}^{ij}_{l+ 2l ^{\prime}}$ for any
$l^{\prime} \in {\Gamma}_0$. Thus instead of the infinite number
of deformation parameters we end up with the finite number. The
twisted deformations ${\m}_l^{ij}$ are in one-to-one
correspondence with the elements of the coset ${\Gamma}_0 / 2
{\Gamma}_0$, which also label the fixed points of the ${\IZ}_2$
action on the four-torus.

For the instanton solutions the resulting equations look a bit
like the self-duality equations with codimension four impurity:
\eqn\kthreeinst{[ Y^i, Y^j ] + {1\over 2} {\ve}_{i^{\prime}
j^{\prime} k k^{\prime}} \sqrt{g} g^{ii^{\prime}} g^{jj^{\prime}}
[Y^k, Y^{k^{\prime}}]  = {\z}^{ij} + \sum_{e \in {\Gamma}_0 / 2
{\Gamma}_0} {\z}^{ij}_{e} \left[ \sum_{l \in {\Gamma}_0} U_{e  + 2
l} \right] {\Omega}} where ${\z}^{ij}$'s are the self-dual
projections of ${\m}^{ij}$'s.

\lref\kraus{P.~Kraus, M.~Shigemori, hep-th/0110035}
\lref\schwarzkonechny{A.~Konechny, A.~Schwarz, hep-th/0107251}

It would be nice to obtain any explicit solution of \kthreeinst\
with ${\z}_e \neq 0$. If ${\z}_e = 0$ then the solutions to
\kthreeinst\ are easy to construct. They correspond to ${\IZ}_2$
equivariant instantons on the noncommutative torus
\schwarzkonechny.

\subsec{Conifold}

\lref\gencon{S.Gubser, N. Nekrasov, S. L. Shatashvili,
hep-th/9811230, \jhep{9905}{1999}{003}}

 To get from the orbifold to the conifold background
we shall mimic the startegy in \gencon. The only difference
compared to the \gencon\ setup is that there one dealt with the
$U(N)$ gauge theory, while here we operate with the $U({\CH})$
gauge fields. But the rest of the story is the same. One starts
with the theory on the orbifold background, turns on the twisted
sector fields, preserving the 8 supercharges (${\CN}=2$ susy in
4d) and then turns on the superpotential giving masses to the
chiral multiplets in the ${\CN}=2$ vector multiplets. For example,
in the ${\IZ}_{k+1}$ orbifold of ${\IC}^2$ case the vacuum
equations will now change to: \eqn\vcm{\eqalign{&  U^{-1} B_{1} U
= {\o} B_{1} \cr & U^{-1} B_{2} U = {\o}^{-1} B_{2} \cr & U^{-1}
{\Phi} U = {\Phi} \cr   [B_i, {\Phi}] = 0, \ i=1,2 \qquad & [ B_1,
B_2 ] = \sum_l \ U^l \left( {\z}_l + m_l {\Phi} \right) \cr}}where
$m_l, {\z}_l$ are the parameters of the generalized $A_k$
conifold.

\newsec{Conclusions and outlook}

The gauge fields/strings duality is a fascinating long-standing
problem\polyakov. We have considered a slightly generalized
version of this duality, which includes noncommutative gauge
fields in various dimensions, and, as a limit, the ordinary gauge
theories. We have defined Yang-Mills algebras and considered
several interesting examples of their representations. These arise
as ${\ap} \to 0$ limits of the algebras of functions on the
D-branes in flat space-time. We have also presented an analysis of
the D-branes in curved backgrounds, namely in those, obtained by
orbifolding from the flat space-time.

The topics left outside of this short note include the
generalizations to the non-trivial $H$-fields (in which case any
universal enveloping algebra of a simple Lie algebra may arise as
an example of Yang-Mills algebra, at least if the dimension of the
latter permits), time-dependent backgrounds (with applications to
cosmology), explicit examples of instantons on the curved spaces,
and, most interestingly, the consequences of the decoupling of the
null-vectors on the closed string side for the open string gauge
invariant quantities \polyakov. We plan to present some of these
considerations elsewhere.

\ndt{\bf Note added.} As the lecture notes were being prepared we
received a manuscript \kraus\ which contained independently
obtained higher-dimensional instanton solutions \newsol, together
with the generalizations involving multiple instantons sitting on
top of each other.

\footatend\vfill\supereject\immediate\closeout\rfile\writestoppt
\baselineskip=14pt\centerline{{\bf References}}\bigskip{\frenchspacing%
\parindent=20pt\escapechar=` \input refs.tmp\vfill\eject}\nonfrenchspacing
\bye